\documentclass{article}
\usepackage[T1]{fontenc}
\usepackage{times}
\usepackage{graphics}

\makeatletter

\newcommand{\LyX}{L\kern-.1667em\lower.25em\hbox{Y}\kern-.125emX\spacefactor1000}

\newcommand{\lyxaddress}[1]{
  \par {\raggedright #1 
  \vspace{1.4em}
  \noindent\par}
}

\makeatother

\begin{document}

\title{The Cochlear Tuning Curve}

\author{Marcelo O. Magnasco}

\maketitle

\lyxaddress{Rockefeller University, 1230 York Avenue, New York \\
marcelo@zahir.rockefeller.edu}

\begin{abstract}
The \emph{tuning curve} of the cochlea measures how large an input is required
to elicit a given output level as a function of the frequency. It is a fundamental
object of auditory theory, for it summarizes how to infer what a sound was on
the basis of the cochlear output. A simple model is presented showing that only
two elements are sufficient for establishing the cochlear tuning curve: a broadly
tuned traveling wave, moving unidirectionally from high to low frequencies,
and a set of mechanosensors poised at the threshold of an oscillatory (Hopf)
instability. These two components suffice to generate the various frequency-response
regimes which are needed for a cochlear tuning curve with a high slope. 
\end{abstract}
Our senses are understood to very different degrees. For example, a fair amount
is known and understood about visual transduction: the eye is like a camera,
with a lens that focuses light onto the sensitive film of the retina. The lens
is solely responsible for the sharpness of imaging, while the rod cells of the
retina are solely responsible for the high sensitivity of the eye; both have
attained fundamental physical limits, diffraction-limited focusing and single-photon
sensitivity. Furthermore we know the molecular players involved in phototransduction
and quite a bit about how these players play together to generate a neural response
to a single photon. By contrast, we know many facts about the ear, but we do
not yet fully understand how to put them together into a coherent picture. We
do know that the ear is nothing like a microphone; if anything, we could say
that the ear is like a camera, with a ``lens'' that sharply focuses sound
(according to frequency) onto a sound-sensitive film. We know that the sharpness
and sensitivity of this process have achieved fundamental physical limitations.
But the intriguing twist of the plot is that, \emph{in the cochlea, the lens
and the film are one and the same}: the hair cells are both the active sound-sensing
elements, as well as the source of the active mechanical feedback of sound that
results in the cochlea's frequency selectivity. If the active mechanisms in
the hair cells are turned off, the cochlea can neither sharply ``focus'' sound
nor amplify faint sounds. This dual role of the hair cells as both lens and
film has made it difficult to unravel various pieces of the puzzle. Furthermore,
the picture is similarly difficult at the molecular level: the transduction
channels which are ultimately responsible for generating an electrical impulse
in response to sound have long eluded identification. First because they are
so few, about tens per hair cell, as opposed to \( >10^{9} \) rhodopsin molecules
per rod cell. Secondly because the high-frequency nature of the detection has
thrust upon these channels dual roles which preclude their direct interaction
with other molecular players which might help in identifying them. Five different
molecules mediate the information cascade between photon detection and the closing
of ion channels that generates an electrical impulse; there is no evidence yet
of any intermediates in hair cells, where the (still hypothetical) tip link-gating
channel complex seems to have simultaneous roles in sound detection, electrical
impulse generation and mechanical feedback. 

The dual roles of the cochlea in amplification and frequency tuning appear to
be two aspects of the same phenomenon. In 1948, T. Gold hypothesized \cite{GOLD}
that in order for the cochlea to provide tuning in view of high viscous damping,
it would have to follow a mechanism similar to regenerative receivers, where
positive feedback can generate at the same time high amplification and sharp
frequency selectivity. This \emph{regenerative hypothesis} lay dormant for many
years, due to what appeared as evidence against it. It slowly resurfaced because
one of its fundamental implications, that misadjusted feedback would result
in sound emission, is to date the only explanation we have for \emph{spontaneous
otoacoustic emissions,} the universally ocurring phenomenon of sound being emitted
by the cochlea. A second line of microscopic experimental evidence for Gold's
hypothesis has been slowly surfacing recently. It was noted in \cite{PNAS},
based on modeling of the transduction channel, that hair bundles appear to be
operating near a Hopf bifurcation, the oscillatory instability behind regeneration,
and experimental confirmation soon followed \cite{MH1,MHJ}. Finally, purely
theoretical considerations provide backing circumstancial evidence: the Hopf
bifurcation posesses intrinsic essential nonlinear characteristics which agree
with otherwise unexplainable nonlinear characteristics of the ear \cite{PRL,TWOTONE}.
It had already been noted by Gold that a feedback loop to keep the system tuned
to the threshold of instability would be required, in the manner of \emph{super}regenerative
receivers; a plausible molecular implementation of self-tuning has been shown
in \cite{CAMALET}; though the exact players are debatable, the mechanism is
generic. 

The problem is that the Hopf instability alone does not seem to explain, by
itself, the mechanical and neural tuning curves of the cochlea. A fundamental
property of sensory systems is that they are ``built'' to be ``used'' by
the brain; i.e., the most illuminating characterization of such systems is not
the ``forward'' transfer responses which tell us how the system reacts to
stimuli, but the ``reverse'' characteristics, in which we try to know what
stimulus caused a given reaction. This is the viewpoint of the brain, which
tries to infer what is out there in the world based on the information coming
out of the sensory systems; an extremely clear example in the H1 neuron of the
fly is given in \cite{SPIKES}, as well as the theoretical infrastructure to
relate these ideas to Bayesian approaches and info theory. In the case of hearing,
this viewpoint was taken very early on; because of the inaccesibility of the
cochlea, first characterizations were of the neural output present in the auditory
nerve rather than mechanical responses, and took the following form. A single
frequency is presented, and its intensity is varied until the output reaches
a specified level---be it spikes per second in the case of the nerve, or a given
amplitude of vibration in the basilar membrane. As the frequency is sweeped,
one reconstructs a \underbar{}curve telling how much input is necessary to elicit
a given response as a function of the frequency, which is called the \emph{tuning
curve} \cite{COCHLEA}; in the case of auditory nerve measurements, this is
called the \emph{neural tuning curve.} The shape of these curve is close to
universal: the tuning curve has one minimum, one specific frequency for which
the intensity required to elicit a given response is minimal, the central frequency
(CF) of the nerve fiber or spot on the cochlea being observed. Around that minimum
there is marked and universal asymetry: a very shallow left flank, and an extremely
steep right flank, and then saturation. The slope of the right flank is a fundamental
measure of auditory acuitiy, reaching the 120 dB per octave in mammals, and
a non-negligible 90-100 db/oct in lower vertebrates such as \emph{Gekko gekko}.
Contrarily, sufficiently close to a Hopf bifurcation systems respond in a universal
way \cite{PRL,CAMALET}, which is symmetric around the resonant frequency, at
least locally. In \cite{PRL} a closed form solution shows the ``tuning curve''
of a single Hopf bifurcation to be locally symmetric; this is computed to be
\( F^{2}=R^{6}+\Delta \omega ^{2}R^{2} \) which is (a) symetric, and (b) shallowly
increasing near the critical frequency. Thus Gold's hypothesis \emph{by itself}
does not fully generate a theory of cochlear function. 

\vspace{0.3cm}
{\centering \resizebox*{0.5\columnwidth}{!}{\includegraphics{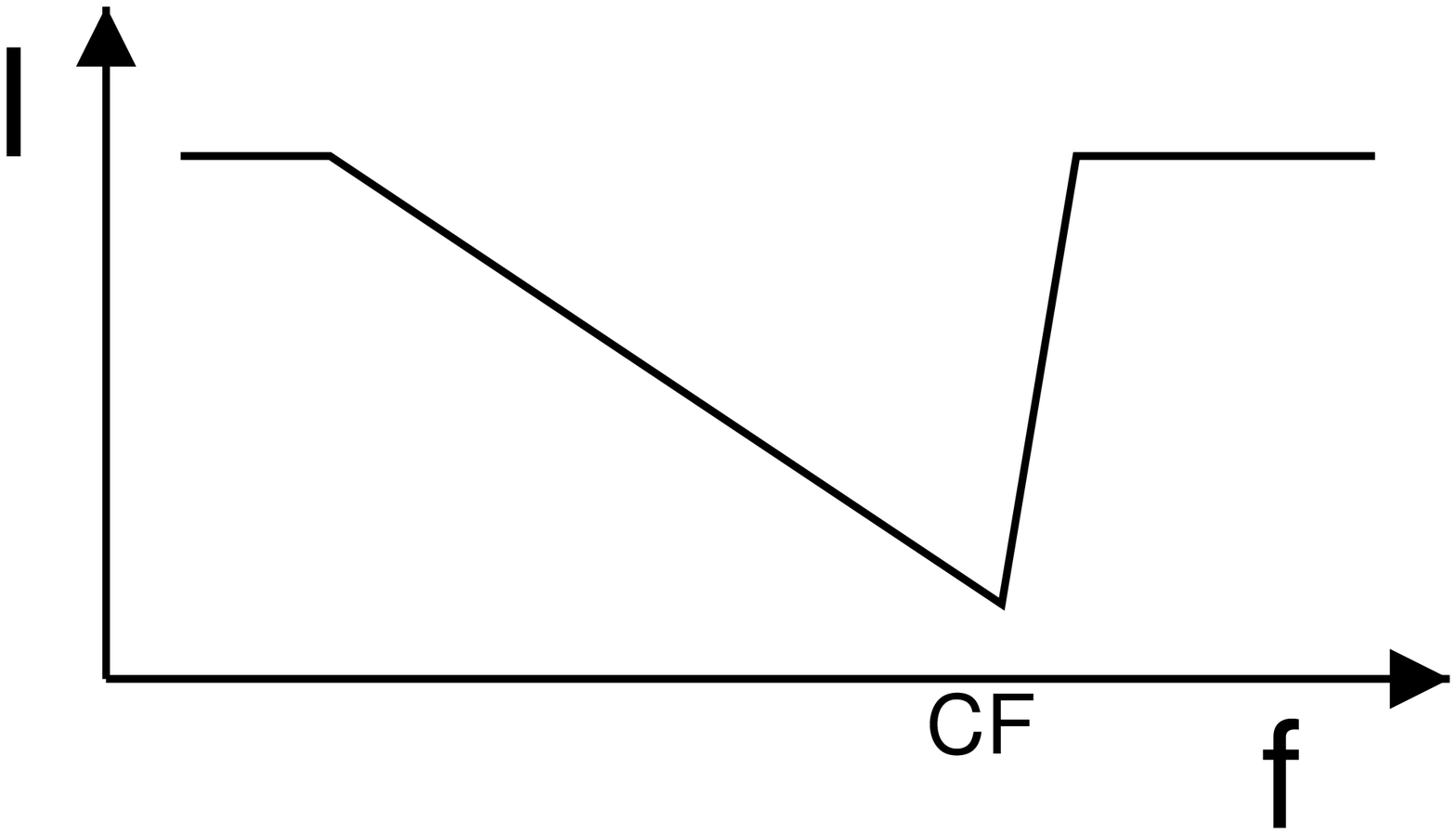}} \par}
\vspace{0.3cm}

\begin{quote}
Figure 1: Generic shape of a tuning curve. 
\end{quote}
\vspace{0.3cm}
{\centering \resizebox*{0.9\columnwidth}{!}{\includegraphics{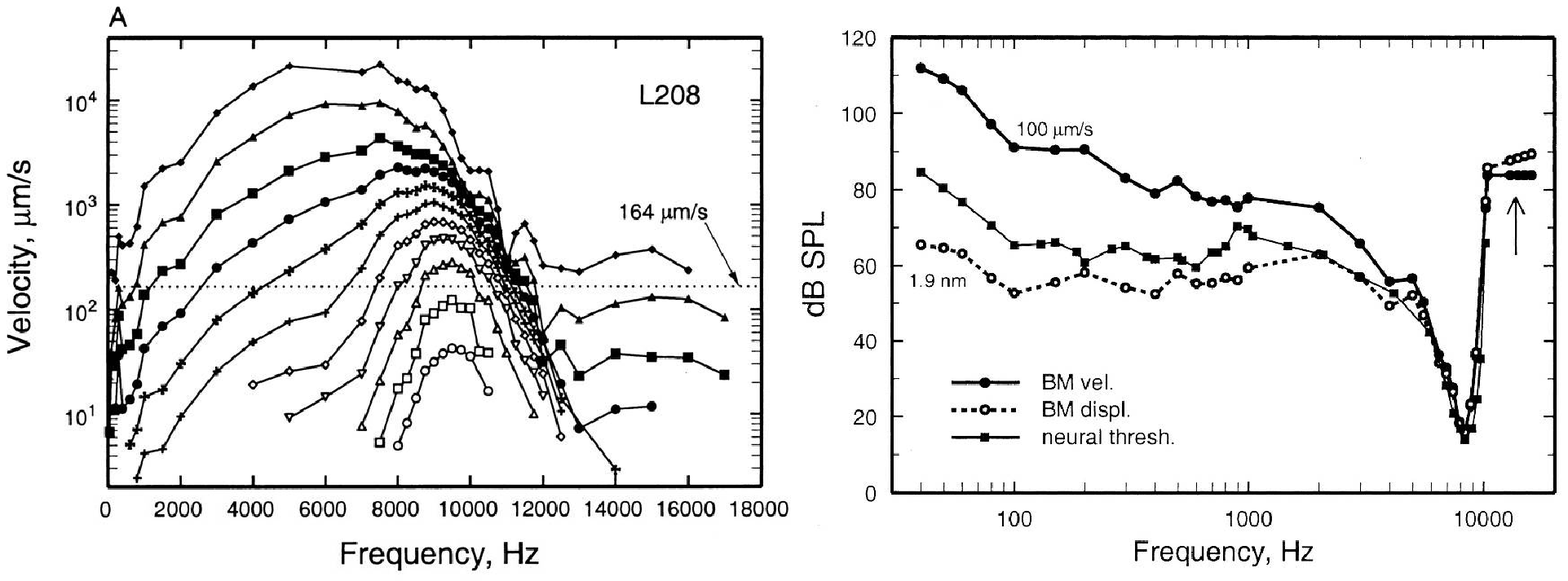}} \par}
\vspace{0.3cm}

\begin{quote}
Figure 2: Cochlear velocimetry data, and its tuning curve, from \cite{RUGG2}.
Courtesy of Mario Ruggero.
\end{quote}
I shall now suply the missing ingredient, the \emph{traveling wave} (TW). In
order to understand tuning curves, one should study the forward quantitative
measurements carefully. Careful examination of cochlear velocimetry data, for
instance from \cite{RUGG1,RUGG2}, show the following. The response curves to
the left of the resonance look quite similar to a Hopf response. However, the
Hopf bifurcation is nonlinearly compressive \emph{only} in the vicinity of the
resonance frequency, while the cochlear velocimetry data is nonlinearly compressive
at the resonant frequency \emph{and all higher frequencies.} The amplitudes
at higher frequencies fall extremely fast, but the responses are still bunched
together at close to \( \frac{1}{3} \)dB per dB. Since the tuning curve is
obtained by intersecting the response curves horizontally and checking which
amplitude intersects the horizontal line at a given frequency, we get immediate
insight: the steepness of the right flank is \emph{not} generated as a result
of high order poles or any such filtering: it is the result of the nonlinear
compression acting only on frequencies greater than or equal to the CF. This
suggests that as sound travels the cochlea, entering at the base, where high
frequencies are mapped, it propagates until reaching the location for its frequency,
where it is nonlinearly amplified. As it continues into the cochlea its amplitude
diminishes rapidly, but it has already been nonlinearly saturated. The unidirectionality
of the traveling wave, together with the fact we have an \emph{array} of Hopf
oscillators rather than only one, give us a strong asymmetry between lower and
higher frequencies; the fact that the traveling wave propagates from high to
low frequencies then dictates that nonlinear saturation occurs in Fig 2 on the
\emph{right} of the resonance. 

\vspace{0.3cm}
{\centering \resizebox*{0.5\columnwidth}{!}{\includegraphics{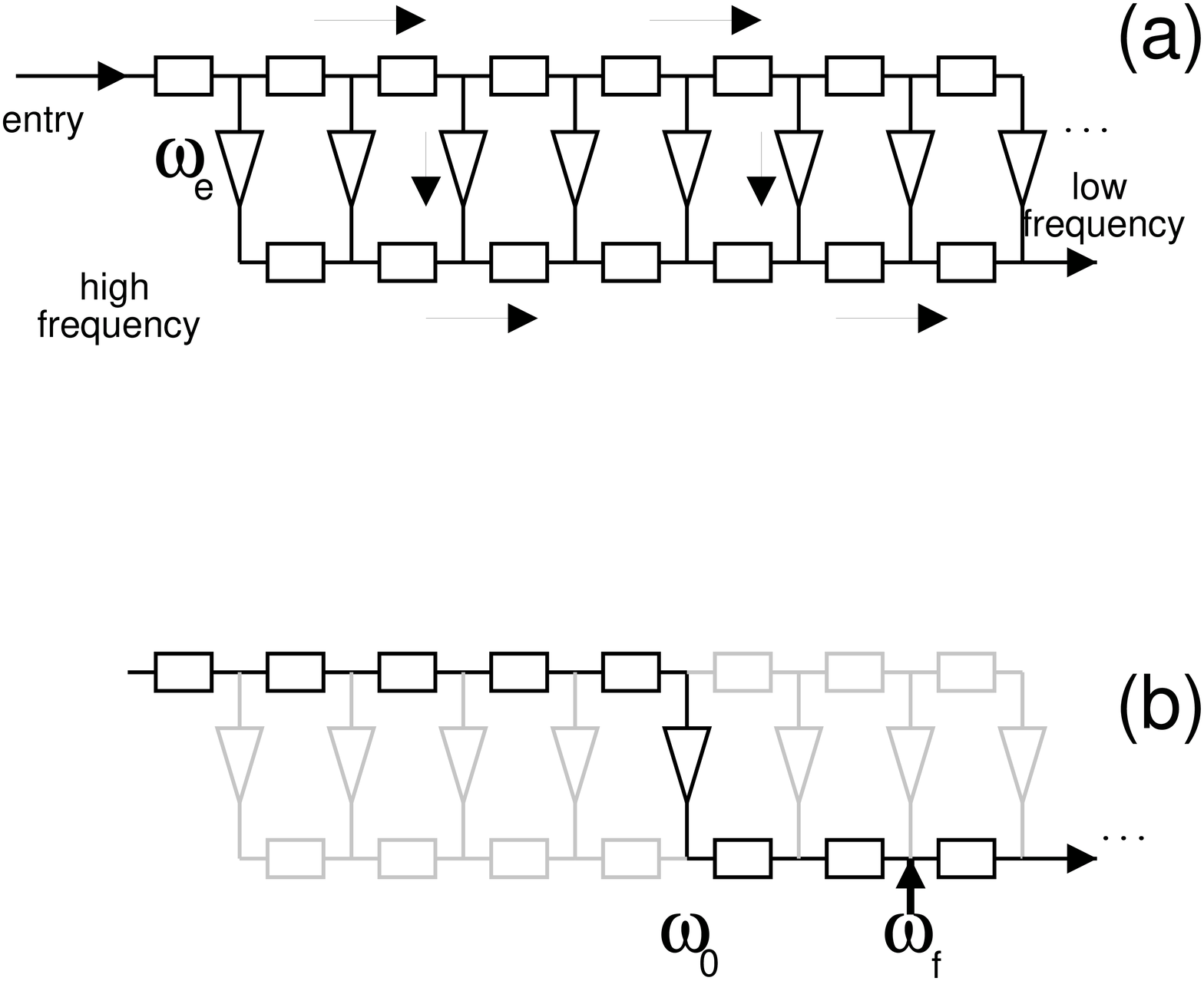}} \par}
\vspace{0.3cm}

\begin{quote}
Figure 3: Scheme of the model. (a) The mechanical equivalent circuit. Rectangles
are passive, linear, weakly tuned elements, while the triangles represent active
elements poised at the Hopf bifurcation. Sound enters through the top left of
the diagram, where high frequencies are mapped, proceeds through the top track
of linear elements, filters down through the active elements, and is collected
through the bottom track of linear elements. The entire structure may be parametrized
by local resonant frequency; highest frequency is \( \omega _{e} \) at the
entry point, and diminishes exponentially as the wave travels. (b) The response
at any given point in the basilar membrane \( \omega _{f} \) is a sum over
all paths of the form shown, which are parametrized by \( \omega _{0} \), and
where \( \omega _{0}>\omega _{f} \): sound moves unidirectionally through this
structure and is not allowed to return back. 
\end{quote}
Are these two elements, Hopf plus traveling wave, enough to explain the tuning
curve? Let us build a simple model of the mechanics of the cochlea. Let's say
that the passive mechanical part of the cochlea is set up as a series of very
weakly tuned low pass filters, organized in a unidirectional series. Let us
use a \( Q=1 \) Lorentzian response for simplicity. Let us say that there are
an array of Hopf oscillators, whose central frequencies track the central frequencies
of the passive part of the system, all arranged in parallel, and then collected
together by a mechanical system identical to the first. (This arrangement has
been selected for analytical tractability; other more complex topologies show
similar results). The response is a sum over all paths through this graph; since
the top and lower portion of the paths are linear they can be explicitly summed. 

In what follows, I shall use specific functions chosen to (a) be reasonable
and (b) be analytically tractable. The qualitative features of the model do
not depend in the least upon this choice. Linear filtering shall be achieved
through a local gain function where \( \omega  \) is the input frequency and
\( \omega _{0} \) the local resonant frequency
\[
g(\omega ,\omega _{0})=\frac{2}{(2\frac{\omega ^{2}}{\omega ^{2}_{0}}-1)^{2}+1}\]
which is equivalent to asserting that locally the filtering being done has a
\( Q \) of 1 at \( 1/\sqrt{2} \) of the local resonant frequency. Then the
overall gain \( G \) is obtained by cascading these filters; we replace the
product of the gains with an exponential of the sum of the logarithms of the
gains, and replace the sum by an integral, to get 
\[
\log G(\omega ,\omega _{0})=\int ^{w_{e}}_{\omega _{0}}\log g(\omega ,\omega _{i})\frac{d\omega _{i}}{\omega _{i}}\]
if we think the filters are distributed on an exponential scale (usual model
for the middle portion of the basilar membrane, since there is an approximately
exponential stiffness); here \( \omega _{e} \) is the entry frequency of the
cochlea, e.g., 20 kHz. Thus, the amplitude \( X \) at the top portion of the
path at a position labelled \( \omega _{0} \) (i.e., CF=\( \omega _{0} \))
as a result of an input with amplitude \( F \) and frequency \( \omega  \)
is 
\[
X_{\omega _{0}}(F,\omega )=F\, G(w,w_{0})\]
and since the integrals can be computed analytically there is a closed form
expression for \( X \); the expression involves second-order polylogs \( Li_{2} \)
and is not particularly illuminating, so we skip it here. Please notice that
the form of \( g \) is arranged so that \( G \) has indeed a maximum at \( \omega _{0} \)
because \( \log g(\omega ,\omega _{0}) \) changes sign at \( \omega _{0} \).
The response \( R \) of a Hopf oscillator forced by amplitude \( X \) at frequency
\( \omega  \) given an internal frequency \( \omega _{0} \) is given by the
root of the cubic 
\[
X^{2}=R^{6}+(\omega -\omega _{0})^{2}R^{2}\]
which is annoyingly solvable: \( A^{3}+dA=B\, \, \rightarrow  \) 
\[
A=\frac{^{3}\sqrt{27B+\sqrt{729B^{2}+108d^{3}}}}{3\, 2^{1/3}}-\frac{2^{1/3}d}{^{3}\sqrt{27B+\sqrt{729B^{2}+108d^{3}}}}\]
 Then the rest of the path to the observation point \( \omega _{f} \) achieves
a gain of 
\[
Y_{\omega _{0}}(\omega ,\omega _{0})=R\exp \int ^{w_{0}}_{\omega _{f}}\log g(\omega ,\omega _{i})\frac{d\omega _{i}}{\omega _{i}}\]
and so the entire contribution of any one path can be analytically computed.
The final response shall be the sum over all paths through intermediate points
\( \omega _{0} \), which obviously depends upon how we weight the density of
local \( \omega _{0} \)s. For simplicity, I shall keep compatibility with the
reasoning above and keep the \( \omega _{0} \) exponentially distributed, thus
the density will equal \( d\log w_{0}=d\omega _{0}/w_{0} \). In addition, we
need to keep track of the relative phases of different paths; the phase lag
for both the linear filter portion and the Hopf elements are all computable
in closed form. Therefore the entire model is solvable in principle as quadratures
of a complex function. 

\vspace{0.3cm}
{\centering \resizebox*{0.49\columnwidth}{!}{\includegraphics{main-10.eps}} 
\resizebox*{0.49\columnwidth}{!}{\includegraphics{main-12.eps}} \par}
\vspace{0.3cm}

\begin{quote}
Figure 4. Maximal path contribution to the final result. In (a) the linear TW
response and the Hopf active elements are ``aligned'', meaning their resonant
frequencies coincide; in (b) they are displaced by a factor of 1.3. CF at observation
point is 1000 Hz. Please note that the response divides into two clear regimes:
\( \omega <\omega _{f} \), to the left of the CF, looks like the Hopf resonance
described in \cite{PRL,CAMALET}, while to the right of the CF, when \( \omega >\omega _{f} \),
all curves are nonlinearly compressed through a cubic root law. Evidently (a)
may not have a sharp tuning curve, for even though the high-frequency regime
is nonlinearly compressed, it impinges on the maximum with zero slope. (b) has
a sharp tuning curve. The implication is that the Hopf elements \emph{need}
to be tuned at higher frequencies than the maximum of the passive TW component. 
\end{quote}
A first, qualitative examination of this model can be achieved by remarking
that the response is going to be dominated by a single path through this graph,
the one with the highest amplitude. For frequencies \( \omega  \) lower than
the CF \( w_{f} \) this path will go through the the top line and then shift
down on the last Hopf oscillator, the one corresponding to the current position.
Thus all lower frequencies than the CF respond exactly like a single Hopf oscillator
composed with a weak linear filter. For frequencies higher than the CF, to lowest
order the dominant path is the one which goes through the Hopf oscillator whose
resonance frequency \( \omega _{o} \) equals that of the input, \( \omega  \);
thus the response, for all frequencies \emph{higher} than the CF, look approximately
like the response at the position whose CF equals the input frequency, composed
with the bottom part of the path, which is a linear attenuation. Thus, to the
right of the CF, all curves stay nonlinearly saturated and just drop down together.
The plot is on Figure 4. Because the response of a Hopf element at \( \omega =\omega _{0} \)
is simply a cubic root, we have a particularly simple expression for the shape
of the response to the right of the resonance: 
\[
Y=F^{\frac{1}{3}}\exp [\int ^{w}_{\omega _{f}}\log g(\omega ,\omega _{i})\frac{d\omega _{i}}{\omega _{i}}+\frac{1}{3}\int ^{w_{e}}_{\omega }\log g(\omega ,\omega _{i})\frac{d\omega _{i}}{\omega _{i}}]\]
where we can see explicitly that the response goes like the cubic root of the
input times a filter. While the formula looks right on paper, a plot of the
response quickly shows not all is well: if the Hopf elements and the TW have
coincident resonances a sharp tuning curve does not obtain, see Figure 4b. 

.

\vspace{0.3cm}
{\centering \resizebox*{0.49\columnwidth}{!}{\includegraphics{tuning.eps}} \par}
\vspace{0.3cm}

\begin{quote}
Figure 5. Tuning curves obtained by intersecting Fig 4b with speed levels 100
and 1000. 
\end{quote}
This picture changes if we take into account all paths, because the broadening
of the Hopf response at higher intensities means that the number of paths that
contribute significantly to the final result changes with intensity. However,
we also need to keep track of relative phases. If phases are not taken into
account at all we obtain a ``wrong'' result: because the width over \( \omega _{0} \)
for which the contribution is significant increases as a \( 2\over 3 \) power
law \cite{PRL}, the nonlinear compression would be utterly obliterated in the
high-frequency regime. There is a trivial solution to this problem (following
the classic ``Cornu spiral'' construction): if the paths have rapidly varying
phases, then only the maxima and the stationary points of phases can contribute.
For simplicity we shall stay with the maximal path contribution only. 

The tuning curve is the inverse of the function computed above with respect
to input and output amplitudes. The slope of this curve can be intuitively understood
as how many lines of the response graph do we intersect as we move horizontally
per unit of frequency change. This number clearly relates to what the slope
of the lines is in the graph, times how many lines do we intersect as we move
\emph{vertically}: the more vertically bunched the lines are, the more bunched
they are horizontally. Thus the immediate effect of the nonlinear saturation
is to triple the number of decibels per octave supplied by the asymptotic front
of the traveling wave on the steep right hand flank of the tuning curve. Furthermore
the bunching and enveloping on the lines on the right hand side do not just
result in a steep right side flank, but also in some invariance of the tuning
curve to the level of response required: as we change our horizontal line up
and down, the tuning curve does not substantially change shape. 

At this point a detailed discussion of the relationship between this (trivial)
model and various well established models in the literature, like Chadwick or
Mammano and Nobile, should ensue, but I feel it's better left for a later occasion.
Also from this model one can verify the assymetry between the two cubic combination
tones (i.e., when \( f_{1}>f_{2} \) the combination tone \( 2f_{1}-f_{2} \)
is stronger than \( 2f_{2}-f_{1} \)). Etc. 

I am deeply indebted to Jim Hudspeth, Victor Martinez-Eguiluz, Boris Shraiman,
Oreste Piro and Bruce Knight for many fruitful discussions about propagation
of sound in the cochlea and its effect on the tuning curves. In particular I
owe to Victor Martinez-Eguiluz the observation that the presence of saturation
on the right hand side of Fig 2 agrees with the traveling wave moving from high
to low frequencies. I am further indebted to Yong Choe, Mark Ospeck, Pascal
Martin and Frank Julicher for discussions on the role of Hopf bifurcations in
auditory detection.

\end{document}